# Properties of high-temperature phase diagram and critical point parameters in silica[†]

Igor Iosilevskiy[1,2]*, Victor Gryaznov[3] and Alexander Solov'ev[2]

[1] Joint Institute for High Temperature RAS, 125412, Izhorskaya st. 13/2, Moscow, Russia
[2] Moscow Institute of Physics and Technology (State University), 141700, Dolgoprudny, Russia
[3] Institute of Problems of Chemical Physics RAS, 142432, Chernogolovka, Russia

Some uncertainties are discussed on the high-temperature phase boundaries and critical point parameters for gas-liquid phase transition in silica ($SiO_2$). The thermal and caloric phase diagrams are compared and examined as being predicted by various theoretical approaches, such as the quasi-chemical representation, the wide-range semi-empirical equation of state and the ionic model under direct molecular dynamic simulation. The theoretical predictions are confronted with handbook recommendations and scanty experimental data on the equilibrium vapor composition over $SiO_2$ boiling. Validity of conventional semi-empirical rules is tested for the theoretically predicted $SiO_2$-phase diagrams. The non-congruence of gas-liquid phase transition in $SiO_2$ is considered for this matter to be used as a modeling body to study the non-congruent evaporation in uranium dioxide and other uranium-bearing fuels at both existing and perspective nuclear reactors.

## 1 INTRODUCTION

High-temperature phase diagrams including the critical point of gas-liquid phase transition as well as thermal ($P, V, T$), caloric ($H, V, T$) and entropic ($S, V, T$) equations of state (EOS-s) for silica ($SiO_2$) are very important in many terrestrial and cosmic applications. Thermodynamic properties of numerous solid modifications of $SiO_2$ are important for physics of the Earth and other planets. Parameters of liquid, gaseous and plasma states of $SiO_2$ are also important in many applications, e.g. natural or artificial bombarding of the Earth and of the Moon (LunarCROSS and Kaguya missions, 2009). In all these cases one meets a strong impact on the planet by a high-speed external space body with subsequent fast transformation of solid $SiO_2$ into gaseous and plasma state during strong shock compression of terrestrial or lunar ground, and then inverse thermodynamic transformation into condensed state via isentropic release of products of huge impact (see e.g. [1] [2] and references therein). Phase transformation of solid $SiO_2$ into gaseous and plasma state occur by fast optical discharge propagation through optical fibers on laser radiation (so-called catastrophic damage (fuse effect) in optical $SiO_2$ fibers (see [3][4] and reference therein).

Besides above mentioned applicative importance of knowledge for gas-liquid phase transition parameters itself, these properties are also important to understand fundamental physical problems. A first example is the general problem of theoretical estimations of critical point parameters for most of metals and compound materials with high critical temperature, which is not easy attainable for accurate experimental exploring. Large uncertainty of such estimations for metallic uranium and some other metals (see e.g. [5] and reference therein) made it important to check theoretical tools on example for $SiO_2$. It should be noted that properties of gas-liquid transition in silica, in particular parameters of its critical point, are poorly known theoretically nowadays. At the same time precise experimental measurement of these properties is very difficult. In particular, there are no experimental data on parameters of high-*T* phase coexistence in silica and its critical point. The main subject of this paper is to compare theoretical predictions for the high-temperature behavior of gas-liquid phase transition in silica and its critical point given by theoretical models. Several models for silica equation of state (EOS) were selected for this purpose. Their predictions were compared with each other and with recommendations of handbooks and scarce experimental data.

It should be emphasized that silica is not a single chemical element but it is a compound combined from two chemical elements. It leads to so-called non-congruence of any phase transition in silica, i.e. to the fact that chemical compositions of both coexisting phases may differ from each other and from a given total composition of two-phase mixture (i.e. total Si/O ratio). Non-congruence is a general phenomenon for phase transformations in any system with two or more "conserved charges", including those in very exotic situations (give examples here [6][7]). Description of properties in non-congruent (or incongruent) phase transitions are more complicated [8][9] than the description of the same properties in simple substance, for example in metals. Features of non-congruent evaporation were studied thoroughly in the high-$T$ uranium-oxygen system, typical product of extreme heating of uranium dioxide in a hypothetical severe accident in nuclear reactor (see e.g. [10] and references therein). The present paper is devoted to the comparison of high-$T$ parameters for simplified (partially equilibrium) variant of phase coexistence in silica, when chemical compositions in liquid and vapor phases are forced to be equal each other. This ordinary scenario of phase coexistence corresponds to well-known Maxwell 'equal square' conditions. Study of the features for non-congruent evaporation in silica is subject of future investigations, which are in progress.

## 2 THEORETICAL MODELS SELECTED FOR COMPARISON

### 2.1 Quasi-chemical model for non-ideal plasma (SAHA-family code).

Selected combination of theoretical models for EOS of $SiO_2$, which is not exhaustive but representative, was chosen for an comparative study of predicted phase diagram of silica in high $T$ – high $P$ region. The first one is the SAHA-model [8][9] – traditional approach in frames of so-called quasi-chemical representation (see e.g. [11][12] etc.). In this approach high-temperature product of $SiO_2$ heating are described as multi-component partially ionized mixture of atoms, molecules, ions and electrons with strong Coulomb interaction, as well as short-range attractive and repulsive forces (i.e. strongly non-ideal plasmas). Minimization of the Helmholtz free energy of this multicomponent mixture is used for calculations of equilibrium composition. Actually it leads to the solution of the system of equations for chemical and ionization equilibrium. A small number of free parameters in this EOS model (in addition to the number of meaningful physical parameters and thermochemical constants like all dissociation and ionization energies of atoms and molecules etc.) was calibrated to reproduce correctly several basic empirical properties of silica, e.g. normal density and heat of sublimation of low-temperature condensed silica. As a result, calculation via SAHA-code [9] gives parameters of gas-liquid phase coexistence (both in congruent and non-congruent scenarios) and additionally equilibrium chemical and ionization composition, as well as the total set of thermodynamic functions in coexisting liquid and vapor phases. In a previous study this approach was applied successfully for calculation of parameters for gas-liquid phase transition in high-temperature uranium dioxide ($UO_{2+x}$), products of extreme heating of nuclear fuel in so-called hypothetical severe accident at nuclear reactor [10]. A special SAHA-code has been created for this purpose [13]. In the present study the SAHA-code was modified for description of the gas-liquid phase transition in $SiO_2$. It should be reminded that in the general case of total equilibrium this phase transition in silica must be non-congruent [9]. Only forced-congruent variant of this phase transition have been calculated in the present study. Such a restriction is justified by great uncertainty of our knowledge for low-$T$ thermodynamic properties of liquid $SiO_2$, which are necessary for adequate calibration of SAHA-code and correct description of parameters for non-congruent evaporation (see below). Finally, parameters of congruent variant of gas-liquid phase transitions in $SiO_2$ were calculated with the use of SAHA-code. It leads to the following thermodynamic parameters of (conventional Van-der-Waals-like) critical point [14].

$$T_c = 6303 \text{ K}; \quad P_c = 3.052 \text{ kbar}; \quad \rho_c = 0.65 \text{ g/cm}^3; \quad (1)$$

$$P_c/(\rho_c R_{SiO2} T_c) = 0.180; \quad S_c/R_{SiO2} = 14.26; \quad \text{Average molecular weight} = 44.87$$

Here $T_c$, $P_c$, $\rho_c$ and $S_c$ denote temperature, pressure, density and entropy of $SiO_2$ in critical point. $R_{SiO2}$ is universal gas constant related to $SiO_2$ molecular weight (0.13837 J/g·K)
These calculations give also following equilibrium vapor composition for concentrations of neutral and charged species $n_i$ (in [cm$^{-3}$]) [15] (see also section 3.4. below):

$$n_{SiO2} = 2.960 \cdot 10+21 \ / \ n_{SiO} = 3.519 \cdot 10+21 \ / \ n_{Si} = 4.344 \cdot 10+18 \ / \ n_O = 8.676 \cdot 10+20 \ / \ n_{Si2} = 3.681 \cdot 10+14$$
$$n_{O2} = 1.330 \cdot 10+21 \ / \ n_{Si(+)} = 4.020 \cdot 10+16 \ / \ n_{O(+)} = 3.409 \cdot 10+14 \ / \ n_{e(-)} = 4.054 \cdot 10+16 \quad (2)$$

## 2.2 Semiempirical wide-range EOS.

The second EOS model, which was selected for comparison [16], is a semi-empirical "wide-range" equation of state. This type of EOS-model is a specially constructed algebraic form with a number of free ("calibrated") parameters. It could be considered as a far improvement of a simple generic Van-der-Waals formula for EOS $P(V,T)$. The presently selected variant of wide-range equation of state was developed earlier by R. More *et al.* [17]. In frames of this model electronic contribution is described by a well-known Thomas-Fermi (TF) model combined with Wigner-Seits average atomic cell approximation. Well-known disadvantages of simple TF model, i.e. noticeable overestimation for critical pressure and temperature, as well as pressures near normal conditions, were corrected by special additional "cold term" (special bonding contribution for electrons and Cowan's model for ion contribution [18]), which was added to the EOS to provide realistic EOS behavior at low temperature and relatively low density. This improved variant of EOS model of D. Young et al. [19] (MPQeos-JWGU) was modified by Stefan Faik et al. [16] [20]. It was applied for calculation of hydrodynamics of quasi-isobaric expansion of thin $SiO_2$ foil under heavy ion beam (HIB) volumetric heating in future perspective experiments within HEDgeHOB collaboration [21] in FAIR Program in GSI (Darmstadt) (see [22] and references therein). Besides that the improved version of MPQEOS was used for calculation of parameters for gas-liquid phase transition in silica. For this purpose the new version of MPQEOS was calibrated at low temperature on known properties of $SiO_2$, as well as it was done for above mentioned SAHA-EOS model. Two basic parameters were used for calibration: $\rho_0$ – normal density of amorphous $SiO_2$ at $T = 0$, and $\Delta_S H^0$ – heat of evaporation in room conditions. The calibrated EOS was validated at $T >> 0$ by comparing thermodynamic parameters of gas-liquid phase transitions, such as $P(T)$ – saturation vapor and $H(T)$ – enthalpy diagram with results of other theoretical EOS models and known data from handbook recommendations (see below). Gas-liquid phase coexistence parameters were calculated via well-known Maxwell "equal square" construction for isotherms $P(V)$ (or 'double-tangent' construction for free energy $F(V,T)$). This modifications lead finally to rather realistic values for the predicted parameters of the critical point and high-$T$ part of gas-liquid phase coexistence. They are compared in Table 1 below with several other variants for critical point parameter estimations within semi-empirical wide-range EOS models (see [23] [2] and references therein)

Table 1. Critical point parameters estimated for $SiO_2$ within semi-empirical wide-range EOS models.

| EOS model | Temperature (K) | Pressure (GPa) | Density (g/cm$^3$) | Reference |
|---|---|---|---|---|
| Faik *et al*. (MPQeos-JWGU) | 4862 | 0.5506 | 0.650 | This work [16] [20] |
| Young and Alder (1971) | 12400 | 0.60 | 0.650 | [24] |
| Ahrens and O'Keefe (1972) | 13500 | 0.642 | 0.637 | [25] |
| Melosh (ANEOS) (2007) | 5398 | 0.189 | 0.549 | [23] |
| Kraus (M-ANEOS) (2012) | 5130 | 0.130 | 0.508 | [2] |

Note that critical temperature and pressure estimated within MPQeos-JWGU and old versions of wide-range EOS [24] and [25] differ noticeably from corresponding critical parameters, predicted by the SAHA-model (see above). It should be pointed out that any phase transitions, which could be calculated in frames of semi-empirical wide-range EOS by using Maxwell construction, are forced-congruent, i.e. coexisting phases have forcedly equal chemical compositions. Besides that the presently used version of semi-empirical wide-range equation of state gives not any information about equilibrium chemical and ionic composition.

## 2.3 Direct numerical simulation (DNS) approaches
2.3.1 Classical ionic model under molecular dynamics simulations

Two such EOS obtained by DNS were selected for a comparative study of phase transition predictions. The first one is the typical approach for description of phase transitions in two-component ionic compounds like NaCl, $SiO_2$, $H_2O$ etc. as classical binary ionic mixture of point-like positive and negative ions, say $Si^{4+}$ and $O^{2-}$ [26][27], with semi-empirical pairwise Coulomb-like repulsive and attractive potentials, corrected at close distances by Lennard-Jones additives for taking into account quantum effects (see [27] and references therein for details). Thermodynamic functions of such two-component system have been calculated within [27] in frames of direct MD simulations. The calculation results for thermal EOS (a great number of points $\{P_i(\rho_i,T_i)\}$) were approximated consequently by a simple algebraic formula ($P(\rho,T)$-EOS envelope) as pressure on density expansion with coefficients depending on temperature. This analytic fit has been used in [27] for standard calculation of parameters for the high-temperature part of the gas-liquid phase coexistence boundaries (binodals), including the critical point. It gave the following parameters for the critical point [27] (compare it with estimations in Table 1):

$$T_c = 11976 \text{ K}; \qquad P_c = 2.0 \text{ kbar}; \qquad \rho_c = 0.58 \text{ g/cm}^3 \qquad (3)$$

It should be noted that the equilibrium system considered in [26][27], composed initially from only *two* basic constituents: ions $Si^{4+}$ and $O^{2-}$ (or ions with effective charges $Si^{1+}$ and $O^{0.5-}$ in [26]), has evident limitation. This model can reproduce in equilibrium vapor phase only a restricted composition of neutral (and charged) bound clusters (molecules): e.g. $SiO_2$, $Si_2O_4$, $Si_3O_6$, ($SiO^{2+}$, $SiO_3^{2-}$)… etc. At the same time it can not reproduce such neutral atoms and molecules as: SiO, Si, O, $O_2$ etc. But it should be stressed that just these species play a major role in equilibrium vapor composition over the boiling $SiO_2$ in the critical point (see Eqn. (2) above) and at lower temperature (see section 3.4. below). As a result of this limitation the considered MD simulation technique [26][27] cannot describe adequately totally equilibrium *non-congruent* phase coexistence, but only a congruent one, i.e. chemical composition in liquid and vapor phases is forcedly equal in fact independently from numerical accuracy of DNS calculations.

2.3.2 Quantum Molecular Dynamics (DFT/MD)

The second *ab initio* approach is the most complicated and contemporary one. It considered high-$T$_high-$P$ products of $SiO_2$ heating as a quantum mixture of nuclei Si and O (or highly ionized ions) in equilibrium with quantum electronic subsystem. Electronic contribution is described via density functional theory, while ionic degrees of freedom are described in frames of Molecular Dynamics procedure (DFT/MD) (see [28][29] and references therein). Simulations for liquid phase only are available presently because DFT/MD calculations for systems like $SiO_2$ are very complicated. Simulations in [28] [29] provided thermal EOS $P(\rho,T)$ along isotherms $T = 3 - 6$ kK and in density range: $\rho > 2.2$ g/cm$^3$, which is close to expected density of boiling $SiO_2$. Equilibrium vapor phase was not simulated in [28] [29]. Only simulation results for the liquid phase were used in present comparisons with predictions of other EOS models. We used the DFT/MD data [28] directly and extrapolated the data from [29] to the limit of zero-pressure in a realistic assumption that at low enough temperature ($T \ll T_c$) zero-pressure isobar coincides practically with the phase boundary of equilibrium condensed phase, i.e. solid or liquid. Results of such extrapolation are exposed at Figure 1 (below).

## 3. COMPARISON OF PARAMETERS FOR THEORETICALLY PREDICTED PHASE DIAGRAMS AND CRITICAL POINTS

### 3.1 Comparison in density–temperature phase diagram

It is well known that density-temperature boundaries of two-phase regions for very many materials obey to two semi-empirical rules (see e.g. [30]):
 - this boundary is always a strongly convex figure, which is totally confined within typical triangle (for details see [5] [31])
 - It obeys to the semi-empirical "rule of rectilinear diameter" of Calliete-Matthias.

Three theoretically predicted phase boundaries are compared in the $\rho-T$ plane (Figure 1). All three EOS expose significant discrepancies in their predictions. It is necessary to discriminate all these different predictions. In our opinion the empirical or *ab initio* calculated density-temperature dependence $\rho(T)$ for liquid $SiO_2$ along zero-pressure isobar, extrapolated in high-$T$ region, can make such a discrimination. One can conclude from Figure 1 that all three EOS should be improved significantly to reach agreement with results of DFT/MD simulations [28] [29] for density along the $P = 0$ curve.

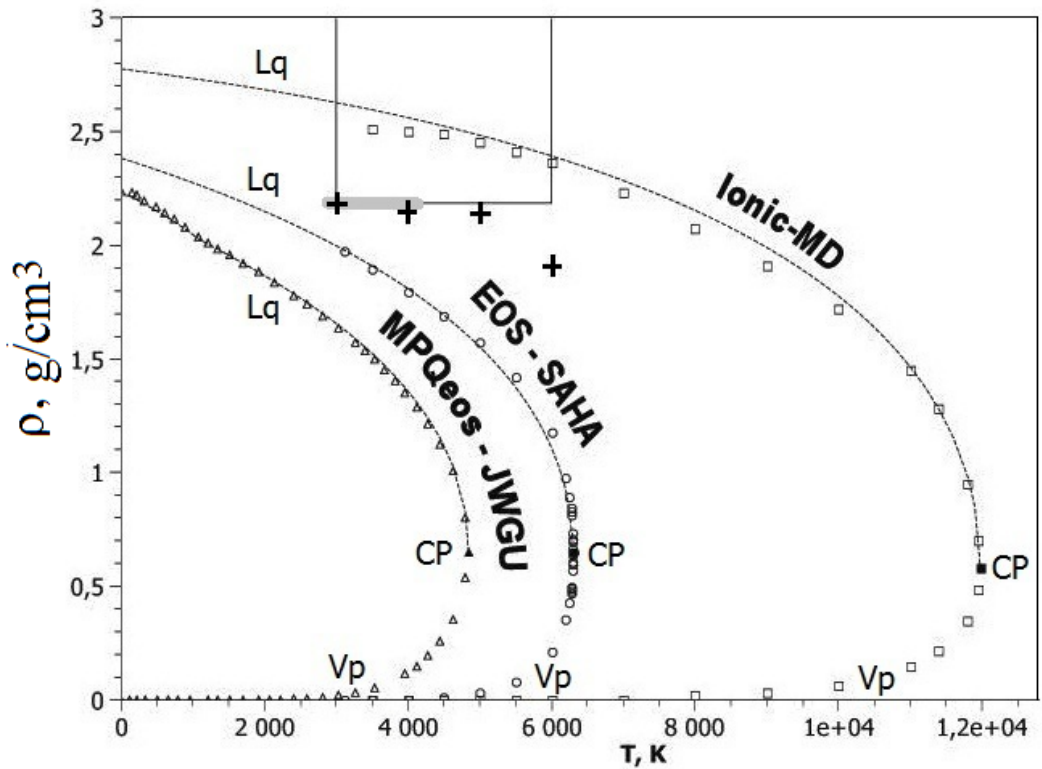

**Figure 1**. Density-temperature ($\rho-T$) phase diagram for silica $SiO_2$. <u>Notations</u>. Δ – MPQeos-JWGU [16][20]; ○ – EOS-SAHA [9]; □ – Ionic model [27]; Lq – liquid; Vp – vapor; CP – critical point; dashed lines – approximation of liquid density by Guggenheim formula (4); box and crosses (+) – area of liquid density calculations $\rho(P)_{T = const}$ via *ab initio* DFT/MD [29] and extrapolation of these data to zero-pressure conditions; thick grey line at $T=3000-4000$ K and $\rho \approx 2.2$ g/cm$^3$ – liquid density at $P = 0$ according to *ab initio* calculations [28].

It is known (see e.g. [32]) that many $\rho-T$ diagrams for gas-liquid phase boundaries surprisingly well obey to a very simple analytical fit – the so-called Guggenheim formula (4).

$$\omega \equiv \frac{\rho(T)-\rho_c}{\rho(T)} \approx \pm B\theta^\beta + C\theta \qquad \left(\theta \equiv \frac{T_c - T}{T_c}\right) \quad (B,C,\beta = const) \tag{5}$$

Here $\rho(T)$ is density of coexisting phases (plus and minus before $B$ correspond to the liquid and vapor phases). $B$, $C$ and $\beta$ are adjusted parameters. This approximation has been performed within this paper for the liquid density, predicted by three EOS models. All three liquid boundaries could be successfully approximated by formula (4), see Figure 1.

### 3.2 Comparison in enthalpy–temperature (caloric) phase diagram

The next comparison is exposed at Figure 2 for caloric EOS, i.e. enthalpy $H(\rho,T)$. Predictions of two tested EOS-s, i.e. SAHA and MPQeos-JWGU, are compared with recommendations of JIHT database IVTANTERMO [33] for enthalpy of coexisting condensed and vapor phases. The first is enthalpy of solid and liquid $SiO_2$ at relatively low temperatures $T < 4500$ K. The second is the enthalpy of hypothetical ideal monomolecular gas of molecules $SiO_2$ (top branch at Figure 2 for $0 < T < 6000$ K). Calculations via SAHA-code were provided for temperature $T \geq 3120$ K, while calculations via MPQeos were provided for $T > 0$. One can see from Figure 2 that the enthalpy of vapor phase given by both comparing EOS are in satisfactory agreement with recommendations of database [33]. At the same time it should be noted that true enthalpy of equilibrium vapors over the boiling and solid $SiO_2$ can not coincide with recommendation [33] for vapors because the latter do not take into account equilibrium dissociation of $SiO_2$ in vapor phase ($SiO_2 \Leftrightarrow SiO + \frac{1}{2}O_2$). It should be stressed that calculations via SAHA-EOS predict high degree of dissociation via this reaction: i.e. degree of dissociation ~ 50% in critical point and even more at 3000 K $< T < T_c$ (see below). Note that these calculations are provided in so-called forced-congruent regime, i.e. by Maxwell rule. At the same time we don't know at present moment equilibrium degree of dissociation in vapors in (totally equilibrium) non-congruent regime of $SiO_2$ evaporation.

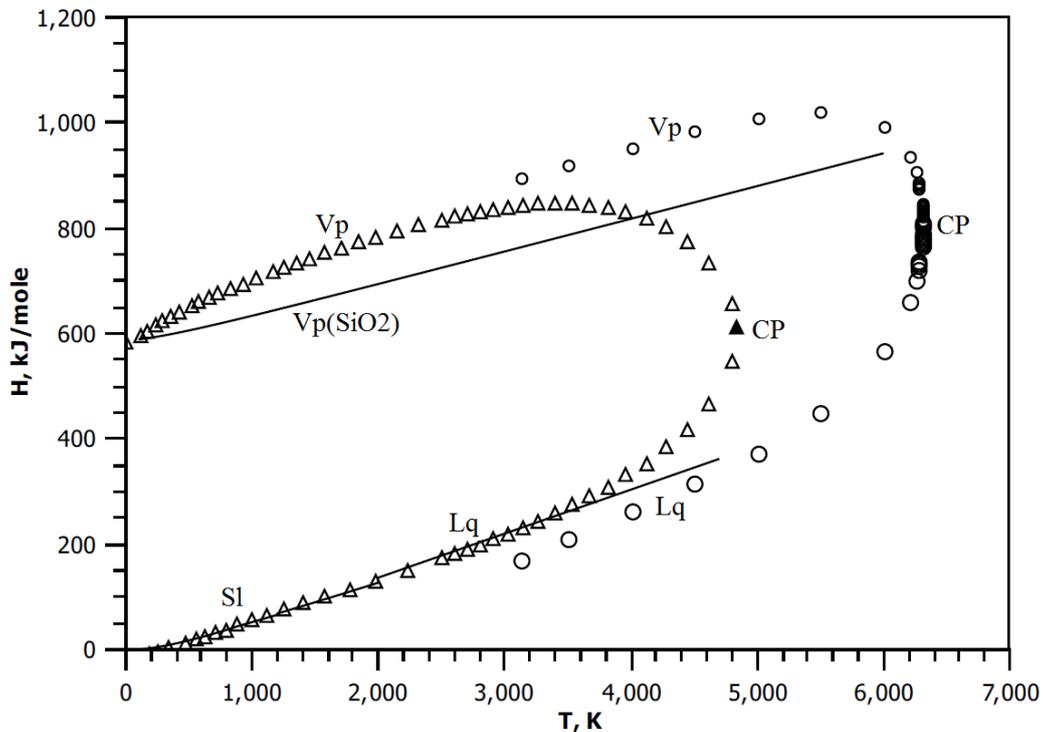

**Figure 2.** Caloric diagram (enthalpy – temperature) of gas-liquid-solid coexistence in silica. *Notations*: Δ – MPQeos-JWGU [16] [20]; ○ – EOS-SAHA [9]: Lq – liquid; Vp – vapor (total composition); CP – critical point; Solid curves – enthalpy of solid (Sl) liquid (Lq) and vapor phases of silica, recommended by IVTANTERMO database [33]: Vp(SiO2) – enthalpy of ideal gas of molecules $SiO_2$ (without dissociation). Small discontinuity in enthalpy of condensed phase between Sl and Lq at $T_{melt} = 1996$ K – heat of melting ($\Delta H = 9.6$ kJ/mole [33]).

## 3.3 Comparison in pressure–temperature phase diagram

Pressure-temperature dependences for many materials obey well to an old semi-empirical rule – the quasi-linear behavior of the saturation curve in Arrhenius coordinates: $\lg(P)$ vs $1/T$. It is known that this rule is strictly valid for low temperature when vapors obey to ideal gas law. At the same time this Arrhenius rule works surprisingly good even for high temperatures close to the critical point. In particular it is valid for three EOS compared in the present paper. It is valid also for predictions of the wide-range EOS-model [23]. The latter were calculated in accordance with recommendation of the NIST/JANAF database. Comparison of all mentioned above $P(T)$ dependences is exposed at Figure 2. All saturation curves have quasi-linear behavior but with different slope and different end-points, i.e. critical points. It should be stressed that only SAHA-code and wide-range EOS-model [23] reproduce a saturation curve (in forced-congruent regime) in satisfactory agreement with predictions of IVTAN database [33]. At the same time should be noted that the data recommended in [33] correspond to Gibbs free energy of condensed $SiO_2$ and therefore correspond to the equilibrium *partial pressure* of an ideal gas of $SiO_2$ molecules over the condensed silica, while the data of the SAHA-code and EOS-model [23] correspond to the equilibrium total pressure of non-ideal vapors over boiling $SiO_2$. It should be stressed (see part 3.4 below) that equilibrium vapors over boiling $SiO_2$ are highly dissociated according to SAHA-code results (see Fig. 4) so that partial pressure of $SiO_2$ in the vapors at $T \ll T_c$ is only a minor component (~ 10 %).

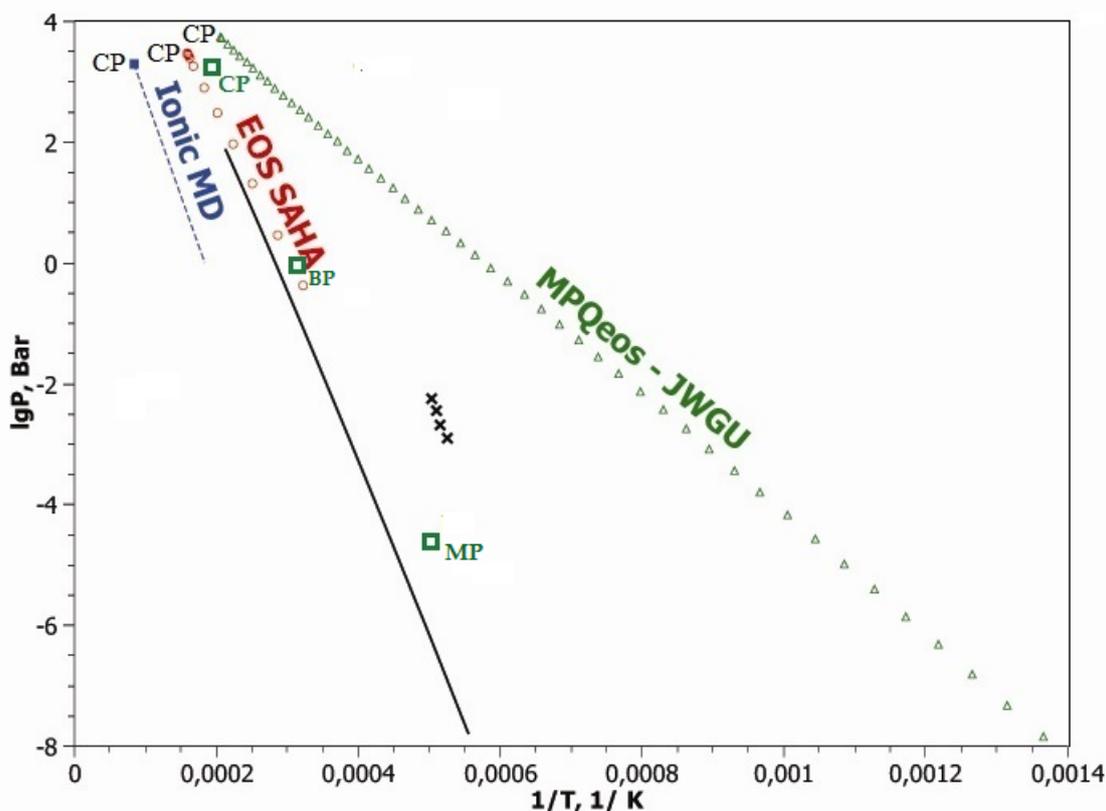

**Figure 3**. Equilibrium vapor pressure over boiling $SiO_2$ in Arrhenius coordinates ($\lg P$ vs $1/T$).
*Notations*: △ – MPQeos-JWGU [16] [20]; ○ – total vapor pressure via EOS-SAHA [9]: CP – critical points; Solid curve – equilibrium partial vapor pressure of ideal gas of $SiO_2$-molecules over condensed silica according to recommendations of IVTAN database [33]; Dashed line – ionic model [27], ■ – corresponding critical point; Open squares □$_{MP}$, □$_{BP}$ and □$_{CP}$ – vapor pressure in melting, boiling and critical points according to the EOS-model [23]. Crosses – experimental data [34].

## 3.4 Equilibrium vapor composition over boiling $SiO_2$
### 3.4.1 Comparison in high-temperature region

Theoretically predicted data about equilibrium composition of the vapor over boiling $SiO_2$ were obtained from calculations of this composition via SAHA model [9][14]. So-called 'individual partition functions' from IVTAN database [33] were used in these calculations for all charged and neutral components of equilibrium Si – O mixture {$SiO_2$, SiO, Si, O, $O_2$, Si(+), O(+),e(-)}. See for example equilibrium composition in critical point (2). These calculations were provided in relatively high temperatures (3000 K < $T$ < 6500 K) and correspond to so-called forced-congruent variant of phase equilibrium (i.e. via Maxwell construction), when chemical compositions of liquid and vapor are forcedly equal to each other and both equal to stoichiometric composition of the $SiO_2$. Unfortunately only scarce experimental data are known presently about equilibrium vapor pressure and vapor composition over $SiO_2$ [34]. Measurements [34] were provided in the vicinity of $T$~ 2000 K.

Both above mentioned experimental and theoretical data are compared in Figure 4. It should be stressed that according to theoretical calculations via SAHA-code [9] the equilibrium Si–O system in saturated vapors over boiling $SiO_2$ is highly dissociated (see Figure 4). The equilibrium fraction of non-dissociated molecules $SiO_2$ is only ~ 10 %, while equilibrium fraction of dissociated products (SiO and O and $O_2$) is about 90 %! It is even more surprising that this strikingly high degree of dissociation just corresponds to existing experimental data [34] at $T$ ~ 2000 K.

It is important that equilibrium vapor composition over boiling $SiO_2$ was calculated also in the wide-range EOS-models [23]. It gives following partial pressures at the boiling point ($T$ = 3177 K and $P$ = 0.1 MPa) [2]:

$$P_{SiO2} = 0.0080 \text{ MPa} \quad / \quad P_{SiO} = 0.0587 \text{ MPa} \quad / \quad P_O = 0.0080 \text{ MPa} \quad / \quad P_{O2} = 0.0253 \text{ MPa} \quad (5)$$

It should be noted that the data for equilibrium composition, calculated via EOS SAHA [9][14] and via EOS-models [23] even in forced-congruent regime proved to deliver surprisingly good agreement with experimental data [34]. The final conclusion about correspondence of theoretical and experimental data on equilibrium vapor pressure over boiling $SiO_2$ could be made after realization of calculation algorithm in totally equilibrium *non-congruent* regime of $SiO_2$ evaporation within SAHA-code [9][10] and extension of these calculations in the region of relatively low temperatures (2000 K < $T$ < 3000 K).

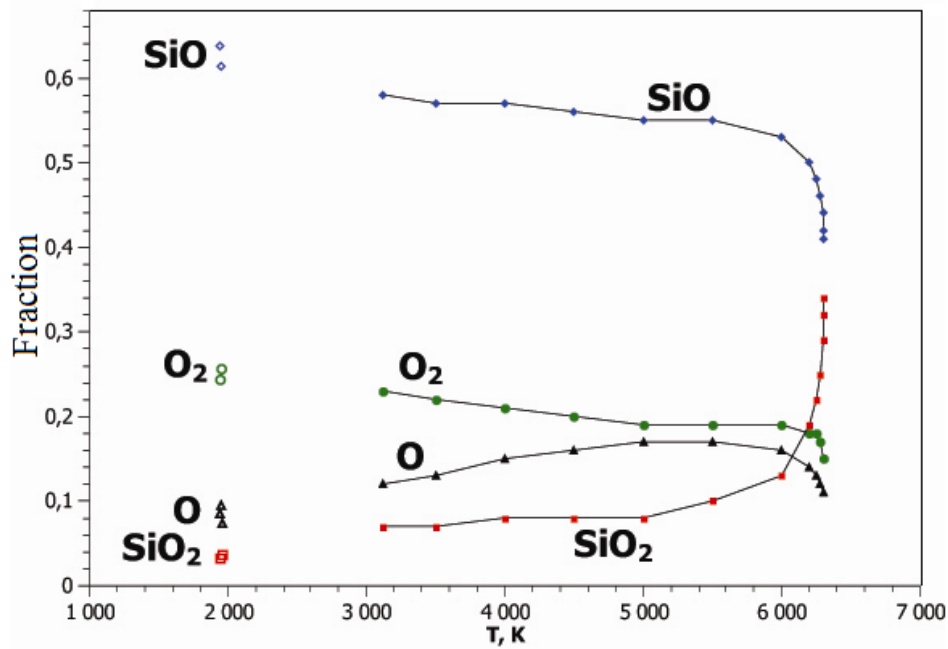

**Figure 4**. Comparison of theoretically predicted molecular and atomic fractions in equilibrium vapor composition over high-temperature boiling $SiO_2$ with experimental data. Filled shapes – calculations at $T$ ~ 3000 – 6500 K via EOS-SAHA [9]; hollow shapes – experimental data at $T$ ~ 1800 – 2000 K [34].

### 3.4.2 Equilibrium vapor composition over condensed SiO$_2$ in zero-temperature limit.

The high degree of dissociation for equilibrium SiO$_2$ vapors, which was revealed in calculations via the SAHA model [10], and its good agreement with scarce experimental data at $T \sim 2000$ K [34] arises an important question: What is the limiting value for equilibrium degree of dissociation in saturated vapors over condensed SiO$_2$ at $T \to 0$? And more generally, what is the equilibrium vapor composition in many two-phase systems, where equilibrium vapors could be more or less associated or decomposed in atoms, molecules (binary (e.g. H$_2$) or multi-component (e.g. UO, UO$_2$, UO$_3$ etc. [9], UF$_n$ etc.), ions and electrons (in single- or even multi-staged ionization), clusters and even equilibrium micro-drops etc.? For ordinary substances, e.g. H$_2$, H$_2$O, CO$_2$ etc. equilibrium vapors in the limit $T \to 0$ are *mono-molecular* (e.g. H$_2$, H$_2$O, CO$_2$ correspondingly). But the question is – how general is this rule, for example, is it valid for the presently considered compound SiO$_2$, and similarly, for the many other extra-important compounds like UO$_2$ [8-10], PuO$_2$, UF$_6$ [35], LiH, NaCl etc.? What is the equilibrium vapor composition (neutral or ionized) for all known metals, including those, which are poorly explored presently?

It should be noted that this question could be answered generally in frames of approach as developed in [36] without a cumbersome free energy minimization procedure via SAHA-code. Only binding energies for all associated complexes as well as basic thermochemical constants for creation of condensed state, such as sublimation and cohesive energies at $T = 0$ (e.g. binding energy of condensed state) are necessary to reveal the discussed composition of equilibrium vapor in the so-called 'vacuum limit' ($T \to 0$ and $\rho \to 0$ at $\mu = const.$ [36]). The main point is that thermodynamics of an equilibrium association mixture (production) is simplified in the vacuum limit. Only a minimal number of species, compatible with electroneutrality condition and given chemical composition dominate in a multi-association system in the vacuum limit. In particular, equilibrium vapor for the case of low-$T$ forced-congruent evaporation in SiO$_2$ may consist in the vacuum limit *only* of SiO$_2$ *or* SiO+½O$_2$. The ruling quantity for correct discrimination of these two cases is the ratio of the handbook values for sublimation energy $\Delta_s H^0$ for reaction (SiO$_2$)$_{solid}$ ⇔ (SiO$_2$)$_{ideal\ gas}$ ($\Delta_s H^0 = 585$ kJ/mole [33]) and dissociation energy $D_{SiO2}$ for reaction SiO$_2$ ⇔ SiO+½O$_2$. ($D_{SiO2} = -220,617$ kJ/mole according to [33]). More correctly, the ruling is the ratio of $\Delta_s H^0$ and the complex $2/3(\Delta_s H^0 + D_{SiO2})$, i.e – 585 kJ/mole vs. – 537 kJ/mole. The lower magnitude of the second value leads to the final conclusion that according to approach [36] the equilibrium vapor over condensed SiO$_2$ (in congruent evaporation) should be fully dissociated in the limit $T \to 0$, i.e consist of SiO+½O$_2$ only. It is promising to consider similar limiting zero-$T$ composition in the case of other compounds, first of all in the case of congruent and non-congruent evaporation in uranium dioxide [9][10]. This work is in progress.

## 5 CONCLUSIONS AND PERSPECTIVES

The current available data are insufficient for high temperature parameters in the SiO$_2$ gas-liquid phase transition, with its critical point parameters indefinite. This is still true even for the simpler scenario of this transition with partial forced-congruent equilibrium. There is nothing to say about the scenario with total non-congruent equilibrium. There are good reasons to expect the anomalous phase behavior due to the non-congruence for silica at high temperature and pressure. The hypothetical non-congruence of phase transitions in silica being taken into account, one should revise ordinary scenarios for all the silica phase transformations in numerous cosmic and terrestrial applications. The parameters of high-temperature gas-liquid phase coexistence for silica predicted by several theoretical approaches have been confronted. The considerable uncertainty has been revealed in these theoretical predictions. In particular, the predicted critical temperature varies in the range of 5000 to 13000 K. It should be stressed also large discrepancy in the predicted caloric phase diagram and in the pressure-temperature dependences as well. Another feature of high-$T$ evaporation in silica is a remarkably high degree of dissociation of SiO$_2$ molecules at equilibrium saturated vapor which is predicted by calculations via SAHA-model. This prediction is validated by its surprisingly good agreement with scarce experimental data. This peculiarity is rather important to comprehend the non-congruent nature of evaporation in SiO$_2$ and other compounds.

Various sub-second experimental approaches look promising to examine the features of phase transition in silica mentioned above. First of all, it is relevant for the impulse laser surface heating (e.g. [37]). It is also relevant for the fast volumetric "quasi-isobaric" heating of porous [38] [39] or stack [40] $SiO_2$-targets by intense heavy-ion-beams (HIB) in frames of HEDgeHOB collaboration [21] [42] within the FAIR Project [41]. A conventional approach with strong shock compression and subsequent isentropic expansion [1][43][2] is quite applicable as well. Concerning this approach, it is intriguing to arrange so-called the "boiling layer" [44] [45] regime of isentropic expansion of shock compressed $SiO_2$-target through the unexplored boundary (binodal) of liquid-gas transition in silica.

From a theoretical view, some variants of so-called *ab initio* approach are desirable for direct numerical simulation ("numerical experiment") at gas-liquid phase transition in $SiO_2$: (Density functional theory, direct quantum Monte Carlo, and Molecular Dynamic simulations, *etc.*).

## 6. ACKNOWLEDGEMENTS

This work was supported by the RAS Scientific Program "Physics of Extreme States of Matter", A.S. was supported by MIPT Education Center "Physics of High Energy Density Matter". I.I. was also partially supported by Extreme Matter Institute – EMMI (Germany).